\documentclass[pdflatex,sn-mathphys,lineno]{sn-jnl}
\usepackage[T1]{fontenc}
\usepackage{amsmath}
\usepackage{braket}

\begin{document}

\title[  ]{Ultra-low-current-density single-layer magnetic Weyl semimetal spin Hall nano-oscillators}

\author*[1,2,3]{\fnm{Lakhan} \sur{Bainsla}}\email{lakhan.bainsla@iitrpr.ac.in}
\equalcont{These authors contributed equally to this work.}

\author*[4]{\fnm{Yuya} \sur{Sakuraba}}\email{SAKURABA.Yuya@nims.go.jp}
\equalcont{These authors contributed equally to this work.}
\author[1]{\fnm{Avinash Kumar} \sur{Chaurasiya}}
\author[1,5,6]{\fnm{Akash} \sur{Kumar}}
\author[4]{\fnm{Keisuke} \sur{Masuda}}
\author[1,5,6]{\fnm{Ahmad A.} \sur{Awad}}
\author[1]{\fnm{Nilamani} \sur{Behera}}
\author[1]{\fnm{Roman} \sur{Khymyn}}

\author[2,7]{\fnm{Saroj Prasad} \sur{Dash}}

\author*[1,5,6]{\fnm{Johan} \sur{Åkerman}}\email{johan.akerman@physics.gu.se}

\affil[1]{\orgdiv{Department of Physics}, \orgname{University of Gothenburg}, \city{Göteborg}, \postcode{41296}, \country{Sweden}}

\affil[2]{\orgdiv{Department of Microtechnology and Nanoscience}, \orgname{Chalmers University of Technology}, \city{Göteborg}, \postcode{41296}, \country{Sweden}}

\affil[3]{\orgdiv{Department of Physics}, \orgname{Indian Institute of Technology Ropar}, \city{Roopnagar}, \postcode{140001}, \country{India}}

\affil[4]{\orgdiv{Research Center for Magnetic and Spintronic Materials}, \orgname{National Institute for Materials Science}, \orgaddress{\street{1-2-1, Sengen}, \city{Tsukuba}, \postcode{305-0047}, \state{Ibaraki}, \country{Japan}}}

\affil[5]{\orgdiv{Center for Science and Innovation in Spintronics}, \orgname{Tohoku University}, \orgaddress{\street{2-1-1 Katahira, Aoba-ku}, \city{Sendai}, \postcode{980-8577}, \country{Japan}}}

\affil[6]{\orgdiv{Research Institute of Electrical Communication}, \orgname{Tohoku University}, \orgaddress{\street{2-1-1 Katahira, Aoba-ku}, \city{Sendai}, \postcode{980-8577}, \country{Japan}}}

\affil[7]{\orgdiv{Graphene Center}, \orgname{Chalmers University of Technology}, \city{Göteborg}, \postcode{41296}, \country{Sweden}}

\abstract{Topological quantum materials  
can exhibit unconventional surface states and anomalous transport properties.  
Still, their applications in spintronic devices are restricted as they require the growth of high-quality thin films with bulk-like properties. 
 Here, we study 10--30 nm thick epitaxial  
ferromagnetic 
Co$_{\rm 2}$MnGa films with high structural order and  
very high values of the anomalous Hall conductivity, $\sigma_{\rm xy}=1.35\times10^{5}$ $\Omega^{-1} m^{-1}$ and the anomalous Hall angle, $\theta_{\rm H}=15.8\%$, both comparable to bulk values 
. We observe a dramatic  
crystalline orientation dependence of the Gilbert damping constant of a factor of two  
and a giant intrinsic spin Hall conductivity, $\mathit{\sigma_{\rm SHC}}=
(6.08\pm 0.02)\times 10^{5}$ ($\hbar/2e$) $\Omega^{-1} m^{-1}$, 
an order of magnitude higher than  
literature values of multilayer Co$_{\rm 2}$MnGa stacks 
\cite{tang2021magnetization, safi2022spin, aoki2023gigantic} and single-layer  
Ni, Co, Fe \cite{wang2019anomalous}, and Ni$_{\rm 80}$Fe$_{\rm 20}$~\cite{wang2019anomalous, seki2021spin}. As a consequence,  
spin-orbit-torque driven auto-oscillations of a 30 nm thick magnetic film are observed for the first time,  
at an ultralow threshold current density of $J_{th}=6.2\times10^{11}$ $Am^{-2}$.  
Theoretical calculations of the 
intrinsic spin Hall conductivity, originating from a strong Berry curvature, corroborate the results and yield values comparable to the experiment. Our results open up for the design of spintronic devices based on single layers of magnetic topological quantum materials. 
}


\maketitle

The nontrivial topology in the electronic band structure of quantum materials makes them potential candidates for emerging technologies such as spintronics~\cite{wolf2001spintronics}, topological electronics\cite{armitage2018weyl, hasan2010coll}, quantum computing~\cite{Sarma2006topo}, and thermoelectrics~\cite{zhou2021seebeck}. Weyl semimetals (WSMs) are one such class of materials, where the nontrivial topological properties arise from  
band touching points, so-called Weyl nodes, in the electronic band structure~\cite{belopolski2019discovery, armitage2018weyl}. WSMs host Weyl fermions in the bulk, as well as surface Fermi arcs connecting Weyl nodes of opposite chirality. Bulk Weyl nodes exist in WSMs that break time-reversal symmetry (TRS)~\cite{armitage2018weyl}, the inversion symmetry (IS)~\cite{armitage2018weyl, soluyanov2015type}, or both~\cite{da2019weyl}. WSMs that break TRS have two Weyl points, while systems that break IS, or both, have four \cite{armitage2018weyl}. WSMs that break TRS are particularly attractive as they allow interplay between magnetism and topology,  
which can result in intriguing quantum states~\cite{armitage2018weyl}. Due to the presence of only two Weyl nodes in magnetic Weyl semimetals, it is easier to observe the chiral anomaly in TRS-breaking WSMs~\cite{armitage2018weyl}, and these materials usually show a large anomalous Hall conductivity (AHC) and anomalous Hall angle due to the large Berry curvature ~\cite{sakai2018giant, belopolski2019discovery, liu2019magnetic}, and therefore have great potential for spintronic applications. 

Recently, different ferromagnetic WSMs have been discovered using angle-resolved photoemission spectroscopy (ARPES), confirming their Weyl nodes and Fermi arc/surface states~\cite{sakai2018giant, belopolski2019discovery, liu2019magnetic,sumida2020spin}. One such material is the Heusler alloy Co$_{\rm 2}$MnGa (CMG), which has a cubic face-centered Bravais lattice, belongs to the space group $Fm3\overline{m}$ ($\#$225), has a room temperature saturation magnetization of ~0.96--0.98~$T$, and stays ferromagnetic up to $\mathit{T_{\rm C}}$ = 690 K~\cite{sakai2018giant,zhou2021seebeck, belopolski2019discovery}.  
CMG single crystals show Weyl lines with drumhead surface states and a large anomalous Hall conductivity (AHC) due to the Berry curvature associated with the Weyl lines. A large value of AHC$\sim$$1.53\times10^{5}$ $\Omega^{-1} m^{-1}$  
was obtained at 2 K for high-quality bulk samples~\cite{belopolski2019discovery}, while a lower value of $1.14\times10^{5}$ $\Omega^{-1} m^{-1}$ was obtained for epitaxial 80 nm thick films~\cite{markou2019thickness}. 

Spin-orbit torque (SOT)~\cite{manchon2019current,shao2021roadmap} has emerged as the most promising and energy-efficient way for switching \cite{fukami2016magnetization, manchon2019current} and excitation of spin waves \cite{fulara2019spin,kumar2024spin}, and the SOT efficiencies of several non-magnetic WSMs have been measured 
~\cite{macneill2017control, shi2019all, bainsla2023large}. Very recently, the SOT efficiency of  
CMG was also studied in  
the CMG/Ti/CoFeB and CMG/Cu/CoTb systems and large negative values of the effective antidamping-like torque, $\xi_{\rm AD}^{\rm eff}$ = 
--0.078 and --0.118, respectively, were obtained~\cite{tang2021magnetization, safi2022spin}. To use SOT from WSMs to manipulate ferromagnets (FM), they need to be deposited in sequence, either directly, as in WTe$_{\rm 2}$/NiFe~\cite{macneill2017control,shi2019all}, or with spacers, as in CMG/Ti/CoFeB~\cite{tang2021magnetization}.  
Both material systems come with the drawback of complicated interfaces, which can perturb or destroy the topological states of the WSMs and hence remove any advantage of using topological materials. Inspired by recent demonstrations of \emph{intrinsic} SOT in single-layer FMs, such as CoPt~\cite{liu2022current} and NiFe~\cite{wang2019anomalous, haidar2019single}, self-induced intrinsic and strongly anisotropic SOT was also very recently reported in $B2$ ordered CMG thin films~\cite{aoki2023gigantic}.   
As the reported values of $\xi_{\rm AD}^{\rm eff}$ are in the range of  --0.15 to 0.20~\cite{aoki2023gigantic, tang2021magnetization, safi2022spin},  
and hence comparable to Pt~\cite{liu2011spin}, the reported CMG films 
do not yet offer any particular advantage for applications. 
It is hence important to obtain high-quality CMG thin films to realize the full potential of its Berry curvature-induced giant SOT for spintronic applications.

Here, we study 10--30 nm thick epitaxial CMG  
films with a much improved structural order that increases with film thickness. From transport measurements on the blanket 30 nm film, we find very high values of the anomalous Hall conductivity, $\sigma_{\rm xy}=1.35\times10^{5}$ $\Omega^{-1} m^{-1}$, and the anomalous Hall angle, $\theta_{\rm H}=15.8\%$, both comparable to bulk values \cite{belopolski2019discovery, sakai2018giant}. Ferromagnetic resonance measurements on the blanket films reveal a high effective magnetization between 0.75 T (10 nm) and 0.9 T (30 nm) and a dramatic (factor of two)  
crystalline orientation dependence of the Gilbert damping constant in the 30 nm film. By patterning the films into Hall bars and microstrips, we extract  
a giant intrinsic spin Hall conductivity, $\mathit{\sigma_{\rm SHC}}=
(6.08\pm 0.02)\times 10^{5}$ ($\hbar/2e$) $\Omega^{-1} m^{-1}$ in the 30 nm film
, which is an order of magnitude higher than 
literature values of multilayer Co$_{\rm 2}$MnGa stacks
\cite{tang2021magnetization, safi2022spin, aoki2023gigantic}, single-layer Ni, Co, Fe \cite{wang2019anomalous}, and Ni$_{\rm 80}$Fe$_{\rm 20}$~\cite{wang2019anomalous, seki2021spin}.
As additional direct experimental proof of the giant intrinsic SOT, we fabricate 150 nm wide nano-constriction spin Hall nano-oscillators (SHNOs)~\cite{demidov2014nanoconstriction,haidar2019single,awad2017natphys,fulara2019spin,zahedinejad2020two,behera2022energy,kumar2023robust,Behera2024} and observe, using micro-Brillouin light scattering ($\mu$-BLS) microscopy~\cite{demokritov2007micro, Sebastian2015}, spin-orbit torque-driven magnetization auto-oscillations of the 30 nm thick single-layer CMG at an ultra-low threshold current density of $J_{th}=6.2\times10^{11}$ $Am^{-2}$.
Theoretical calculations of the  
intrinsic spin Hall conductivity, originating from a strong Berry curvature, corroborate the results and yield values comparable to the experiment.

\begin{figure*}[ht!]
\centering
\includegraphics[width=11.9cm]{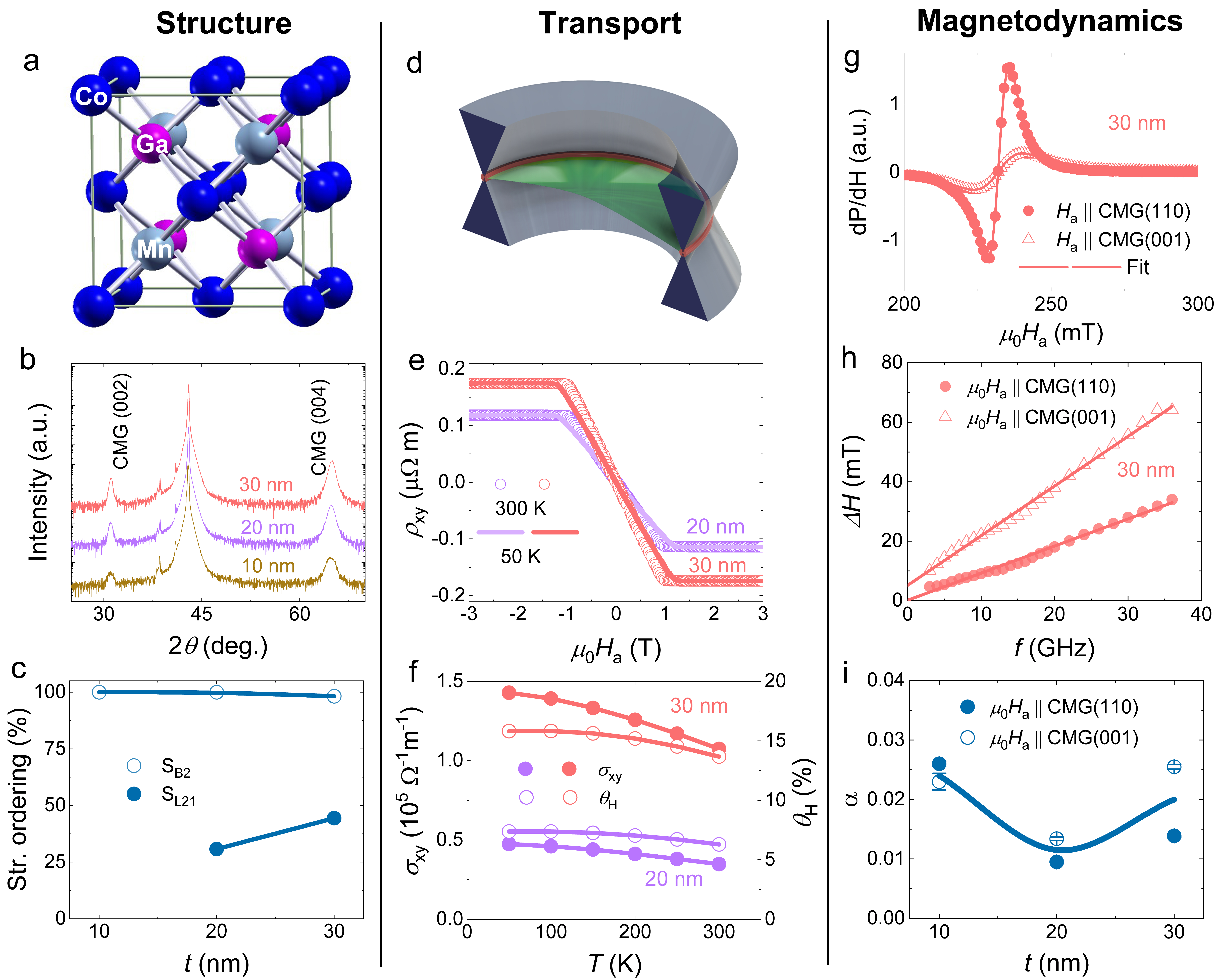}
\caption{\textbf{Growth of high-quality Co$_{\rm 2}$MnGa (CMG) thin films with strong anisotropic Gilbert damping and giant anomalous Hall conductivity.}
\textbf{a}, The crystal structure of $L$2$_{\rm 1}$-ordered Co$_2$MnGa Heusler alloy.
\textbf{b}, Out-of-plane XRD patterns for CMG films with three different thicknesses. \textbf{c}, Structural ordering parameters $S_{\rm B2}$ and $S_{\rm L21}$ \emph{vs.}~film thickness; lines are guides to the eye. \textbf{d}, Schematic of the band structure showing the presence of nodal line and drumhead surface states. \textbf{e}, Hall resistivity, $\rho_{\rm xy}$, \emph{vs.}~applied magnetic field, $\mu_{\rm 0}H_{\rm a}$, at 50 and 300 K for the 
20 and 30 nm films. \textbf{f}, Temperature-dependent anomalous Hall conductivity, $\sigma_{\rm xy}$, and anomalous Hall angle, $\theta_{\rm H}$, for the 20 and 30 nm films;  
lines are guides to the eye. \textbf{g}, Ferromagnetic resonance spectra for 30 nm film.
\textbf{h}, Ferromagnetic resonance linewidth, $\Delta H$, \textit{vs.} \textit{f} for CMG 30 nm film. \textbf{i}, Gilbert damping constant as a function of CMG thickness. \textbf{g}, \textbf{h}, \textbf{i}, Solid and open symbols represent the experimental data points when applied magnetic field, $\mu_{\rm 0} H_{\rm a}$, is parallel to CMG (110) and (001) planes, respectively. \textbf{g}, \textbf{ h}, solid lines are fit to the experimental data \cite{bainsla2022ultrathin}, while solid lines in \textbf{i}, are just guide to eye.} 
\label{Fig.1} 
\end{figure*}

\section*{Results}\label{sec1}

The out-of-plane $2\theta$-$\theta$ x-ray diffraction (XRD) measurements were performed on CMG films with different thicknesses, as shown in Fig. \ref{Fig.1}a. The (002) superlattice peaks and (004) fundamental peaks are clearly visible in the XRD patterns for all the samples, indicating that the films were grown with (001) crystalline orientation and high $B2$ structural ordering ($B2$ refers to Mn-Si order). Further, the epitaxial growth of the films was confirmed by performing $\phi$ scans for the CMG (220) fundamental peak. A clear 4-fold rotation symmetry confirms the epitaxial growth for all films (see supplementary Fig.~1a). $2\theta$-$\theta$ scans were also obtained for the CMG (111) superlattice peak, confirming a $L2_{\rm 1}$ structural order ($L2_{\rm 1}$ refers to Co-Mn order). The (111) peak is present only for the 20 and 30-nm films, indicating the absence of $L2_{\rm 1}$ ordering in the 10-nm film (see supplementary Fig.~1b). The $B2$ and $L2_{\rm 1}$ structural ordering parameters $S_{\rm B2}$ and $S_{\rm L21}$ were estimated using the experimental and calculated XRD intensity ratios~\cite{bainsla2017structural}, and the estimated values are plotted in Fig.~\ref{Fig.1}b. The 30-nm film shows a slightly higher value of $S_{\rm L21}$ compared to the 20-nm film. The obtained values of $S_{\rm B2}$ and $S_{\rm L21}$ confirm full $B2$ and partial $L2_{\rm 1}$ ordering, substantially better than earlier thin film reports on this material \cite{tang2021magnetization, aoki2023gigantic, safi2022spin}.  

Vibrating sample magnetometer measurements showed that the films have an in-plane easy axis and very low coercivity, as shown in the supplementary Fig.~2. The highest saturation magnetization of $4\pi M_{\rm S}$ = 0.84 $\pm$ 0.02 $T$ was obtained for the 30 nm film, which is comparable to earlier reports on thin films
~\cite{tang2021magnetization, safi2022spin}, but less than the bulk values, $4\pi M_{\rm S}$~=~0.96--0.98~$T$~\cite{belopolski2019discovery, manna2018colossal}. Broadband ferromagnetic resonance (FMR) measurements were performed in the frequency range $f= 3-38$ GHz  
with the magnetic field, $H_{\rm a}$, applied parallel to the CMG (110) and (001) planes, as given in Fig. \ref{Fig.1}g for 30 nm film.  
The resonance field ($H_{\rm R}$) and the linewidth ($\Delta H$) were extracted by fitting the experimental data to a sum of symmetric and anti-symmetric Lorentzian derivatives~\cite{bainsla2022ultrathin}.  
The effective magnetization, $\mu_{\rm 0}M_{\rm eff}$, was obtained from fits of \textit{f}~\textit{vs.}~$H_{\rm R}$ to Kittel's equation \cite{kittel1948theory, bainsla2022ultrathin}; $\mu_{\rm 0}M_{\rm eff}$ ranges from 0.75 to 0.9 T for the three films (see the supplementary Fig. 3). 
The effective Gilbert damping constant, $\alpha$, was obtained from linear fits of $\Delta H$ \textit{vs.} \textit{f} to $\Delta H=\Delta H_{\rm 0}+(4\pi \alpha f )/\gamma \mu_{\rm 0}$ (Fig. \ref{Fig.1}h).
The obtained $\alpha$ values are slightly higher than previously reported values~\cite{guillemard2019ultralow}, but show a very interesting hitherto unexplored crystalline orientation dependence, as $\alpha_{001}=$~0.025$\pm$0.001 ($H_a$ applied along the (001) plane) is almost twice as large as $\alpha_{110}=$~0.014$\pm$0.001 ($H_a$ applied along the (110) plane) for the 30 nm film; see Fig. \ref{Fig.1}i. 

AHC values of about $0.45\times10^{5}$ and $1.35\times10^{5}$ $\Omega^{-1} m^{-1}$ were obtained at a temperature of 50 K for the 20 and 30-nm CMG films, as shown Fig.~\ref{Fig.1}e and \ref{Fig.1}f, respectively. These values are higher than those reported for thick ($\sim$80 nm; 1138 S/cm at 2 K) CMG films~\cite{markou2019thickness} and comparable to the bulk case~\cite{belopolski2019discovery}. The origin of the large AHC was discussed in an earlier work by comparing measured angle-resolved photoemission spectroscopy (ARPES) data to the Berry curvature obtained from density functional theory (DFT) calculations and was attributed to the presence of a large berry curvature associated with Weyl lines in the CMG electronic band structure~\cite{belopolski2019discovery,sumida2020spin}. Our high value of AHC for the 30 nm film suggests the presence of Weyl lines and corresponding Berry curvature. The anomalous Hall angle, $\theta_{\rm H}=\sigma_{\rm xy}^{AHE}/\sigma_{\rm xx}$, which reflects the ability of a material to deviate the electron flow from the direction of the longitudinal electric field, was estimated using transport measurements (see the supplementary Fig.~4 for electrical conductivity, $\sigma_{\rm xx}$, measurements), and a record high-value of $\theta_{\rm H}=15.8\%$ was obtained for the 30 nm film.

\begin{figure*}[htt!]
\centering
\includegraphics[width=11.9cm]{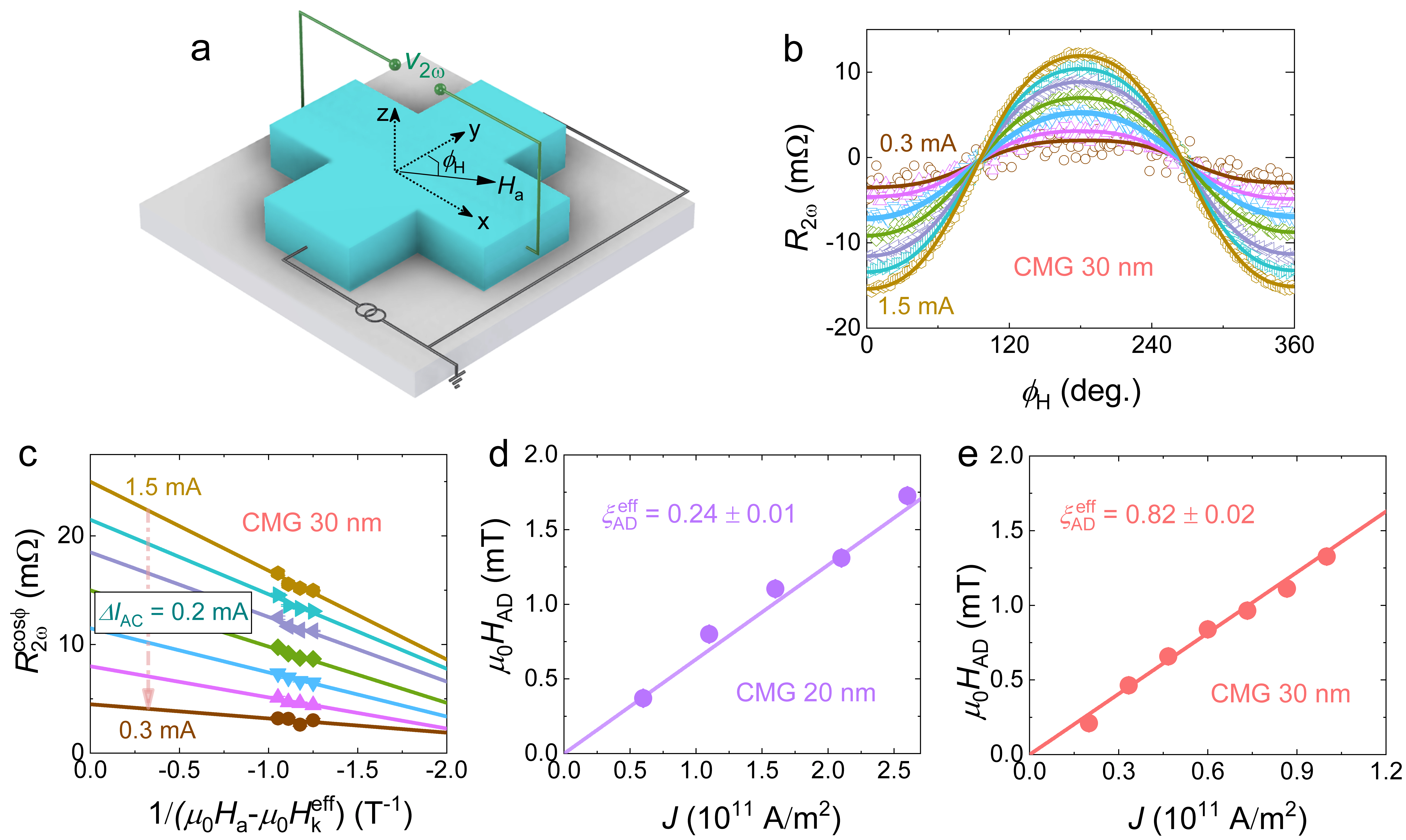}
\caption{\textbf{Second harmonic Hall measurements of the giant intrinsic spin-orbit torque.}
\textbf{a}, Schematic of the measurement set up showing the directions of alternating current, $I_{AC}$, and applied magnetic field, $H_{a}$ with coordinates.   
\textbf{b}, Second harmonic Hall resistance, $R_{\rm 2\omega}$, of the 30 nm CMG film \emph{vs.}~in-plane magnetic field angle, $\phi_{\rm H}$, in  
an applied magnetic field $\mu_{\rm 0}H_{\rm a}=$ 0.2 T, for seven different alternating current values 
$I_{\rm AC}=$ 0.3--1.5 mA, in steps of 0.2 mA. Open symbols are the experimental data points; solid lines are fits to  
Eq.~\ref{eq1}. \textbf{c}, The $\cos\phi$ contribution to $R_{\rm 2\omega}$ ($R_{\rm 2\omega}^{\cos\phi}$)   
as a function of $1/(\mu_{\rm 0}H_{\rm a}-\mu_{\rm 0}\mathit{H_{\rm k}^{\rm eff}})$ at the seven different current values. 
Filled symbols are the experimental data points; solid lines are linear fits. \textbf{d}, \textbf{e},  
The anti-damping-like field $\mu_{\rm 0}\mathit{H_{\rm AD}}$  
\emph{vs.}~current density $J$ for 20, and 30 nm films, respectively. Solid symbols are the experimental data points; solid lines are linear fits.  
The extracted effective anti-damping-like torque value, $\xi_{\rm AD}^{\rm eff}$, is shown in the figures.}
\label{Fig. 2nd harmonic} 
\end{figure*}

Harmonic Hall measurements were performed on the 20 and 30 nm CMG films, as shown in the schematic  
in Fig.\ref{Fig. 2nd harmonic}a. The SOT efficiency was estimated using fits of the second harmonic ($R_{\rm 2\omega}$) Hall resistance \textit{vs.} $\phi_{\rm H}$,  
to~\cite{avci2014interplay, takeuchi2018spin}
\begin{multline}
    R_{\rm 2\omega}=-(R_{\rm AHE}\frac{H_{\rm AD}}{H_{\rm a}-H_{\rm k}^{\rm eff}}+R_{\rm T})\cos\phi_{\rm H}\\+2R_{\rm PHE}\frac{H_{\rm FL}+H_{\rm Oe}}{H_{\rm a}}(2\cos^{3}\phi_{\rm H}-\cos\phi_{\rm H})
\label{eq1}
\end{multline}
where $R_{\rm AHE}$ and $R_{\rm PHE}$ are the anomalous and planar Hall resistances, $H_{\rm FL}$, $H_{\rm AD}$, $H_{\rm Oe}$, and $H_{\rm a}$ are the effective field of the field-like torques, the effective field of the antidamping-like torques, the current induced Oersted field, and the applied magnetic field, respectively. $R_{\rm T}$ is the second harmonic Hall resistance signal due to the thermo-electric effects including the anomalous Nerst effect and spin Seebeck effect. The measured $R_{\rm 2\omega}$ \textit{vs.} $\phi_{\rm H}$ was fitted with Eq.~\ref{eq1} and separated into $cos\phi_{\rm H}$ and $2cos^{3}\phi_{\rm H}-cos\phi_{\rm H}$ contributions. The $cos\phi_{\rm H}$ contribution of $R_{\rm 2\omega}$ ($R_{\rm 2\omega}^{cos\phi}$) \textit{vs.} $1/(\mu_{\rm 0}H_{\rm a}-\mu_{\rm 0}H_{\rm k}^{\rm eff})$ for the 30 nm CMG film is shown in Fig. \ref{Fig. 2nd harmonic}c; $H_{\rm AD}$ and $R_{\rm T}$ values were obtained from the slope and y-intercept of the linear fit of the data, respectively. We assume no net Oersted field contribution in our case, as we are working with a single CMG layer. Figure~\ref{Fig. 2nd harmonic}d\&e show the obtained $H_{\rm AD}$ \textit{vs.} $J$ (current density) for the 20 and 30 nm films, respectively. The slope of $H_{\rm AD}$ \textit{vs.} $J$ was obtained from the linear fit and used to evaluate the effective antidamping-like torque efficiency, $\xi_{\rm AD}^{\rm eff}$, with the relation~\cite{takeuchi2018spin}, 

\begin{equation}
   \xi_{\rm AD}^{\rm eff} = \frac{{2e}{m_{\rm S}}}{\hbar}\frac{\delta{\mu_{\rm 0}H_{\rm AD}}}{\delta J}
\label{eq2}
\end{equation}
\\
where, $\textit{e}$ is the elementary charge, $\hbar$ is the reduced Plack's constant, and $m_{\rm S}$ is the saturation magnetization per unit area. $\xi_{\rm AD}^{\rm eff}$ values of 0.24$\pm$0.01 and 0.82$\pm$0.02 are obtained for the 20 and 30 nm films, respectively (see the supplementary Fig. 5 for CMG 20 nm data). The obtained value of $\xi_{\rm AD}^{\rm eff}=0.82\pm0.02$ is an order of magnitude higher than the previously reported values in systems such as CMG/Ti/CoFeB as mentioned in the introduction~\cite{tang2021magnetization, safi2022spin} and about 5 times the value reported for single layer (110) oriented CMG films~\cite{aoki2023gigantic}. The effective spin Hall conductivity, $\sigma_{\rm SHC}=\sigma_{\rm xx}\xi_{\rm AD}^{\rm eff}=(1.24 \pm 0.01)\times 10^{5}$ and $(6.08 \pm 0.02)\times 10^{5}$ ($\hbar/2e$) $\Omega^{-1} m^{-1}$ were estimated for the 20 and 30 nm CMG films, respectively. The $\sigma_{\rm SHC}$ value for the 30 nm CMG film is at least an order of magnitude higher than reported values for other single layer magnets such as Ni$_{\rm 80}$Fe$_{\rm 20}$~\cite{wang2019anomalous, seki2021spin, haidar2019single}, Ni, Fe, and Co~\cite{wang2019anomalous}.

\begin{figure*}[htb!]
\centering
\includegraphics[width=11.9cm]{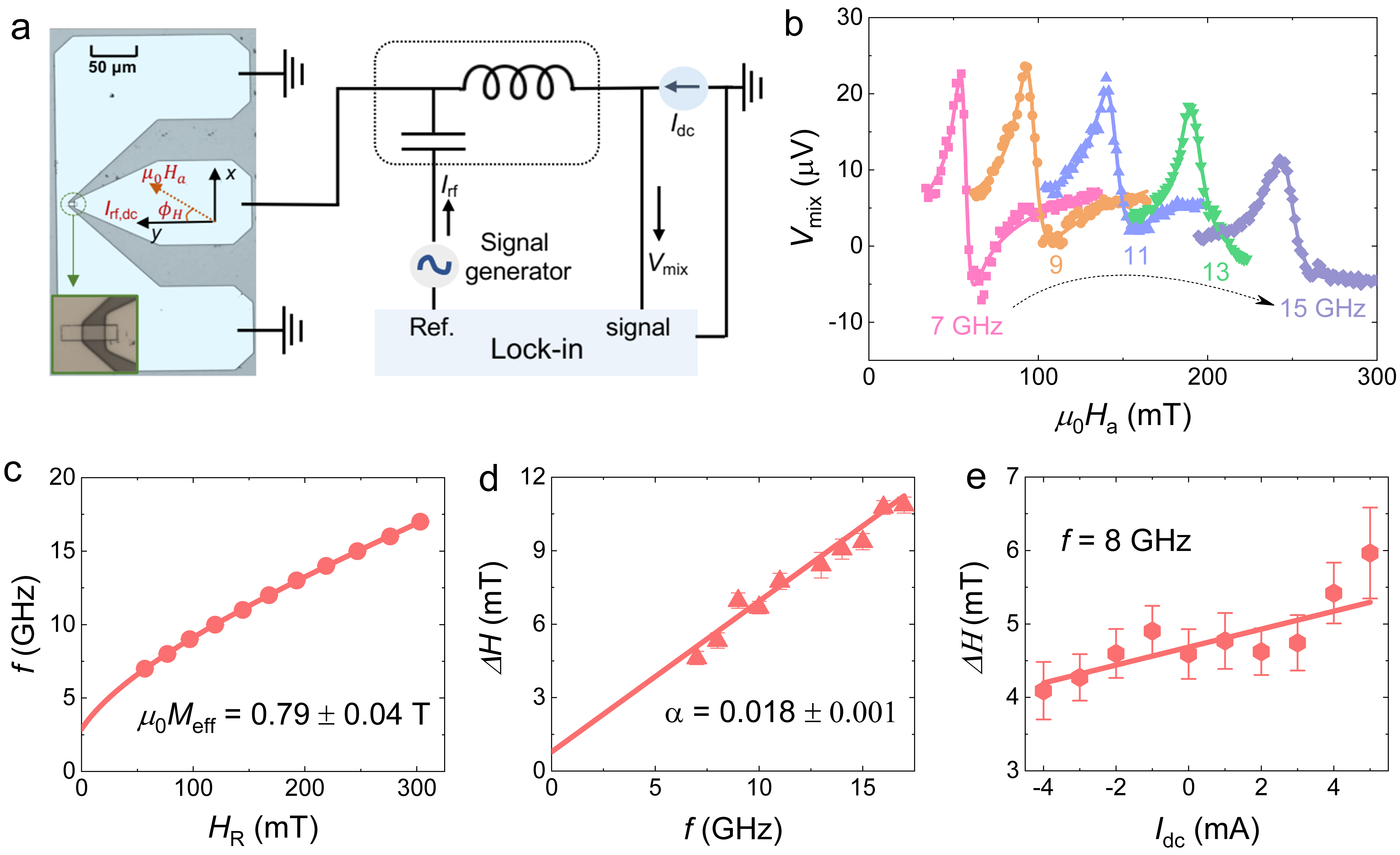}
\caption{\textbf{ST-FMR measurements of the giant intrinsic spin-orbit torque.
}
\textbf{a}, Schematic of the ST-FMR measurement setup. \textbf{b}, Five representative STFMR curves in the frequency range $f=$ 7--15 GHz. Solid symbols are the experimental data; solid lines are fit to, $V_{\rm mix}=SF_{\rm S}(H_{\rm a})+AF_{\rm A}(H_{\rm a})$  
~\cite{demasius2016enhanced, liu2011spin}. \textbf{c}, \textbf{d}, Frequency \textit{f} \textit{vs.} resonance field $H_{\rm R}$ and resonance linewidth $\Delta H$ \textit{vs.} frequency \textit{f}, respectively. Here, the solid symbols show the values obtained by fitting the V$_{\rm mix}$ signal, and the solid lines show the fit to the data. The obtained values of effective magnetization, $\mu_{\rm 0}M_{\rm eff}$ and the Gilbert damping constant $\alpha$ are shown in Fig. \ref{fig. stfmr}c and \ref{fig. stfmr}d, respectively. \textbf{e}, Resonance linewidth $\Delta H$ \textit{vs.} dc bias $I_{\rm dc}$ measured at a frequency of 8 GHz, here solid symbols are the data points obtained by fitting the V$_{\rm mix}$ signal and the solid line is a linear fit to the obtained data. \textbf{b}-\textbf{e}, All the measurements are done with $\phi_{\rm H}=60^{\circ}$.}  

\label{fig. stfmr} 
\end{figure*}

To further estimate the SOT efficiency, spin-torque ferromagnetic resonance (ST-FMR) measurements were performed on rectangular 4\,$\times$14 $\mu m^2$ microstrips fabricated with the longer axis along the CMG(110) plane (measurement schematic is given in Fig.~\ref{fig. stfmr}a),  
as a recent study found the largest SOT efficiency when the current flows along the CMG (110) plane~\cite{aoki2023gigantic}, and using FMR measurements on blanket films we found that $\alpha$ is minimum when we measure the films with $H_{\rm a}$ along the CMG (110) plane. Anisotropic magnetoreristance (AMR) measurements were performed on the ST-FMR microstrips  
and negative AMR values in the range from --0.37 to --0.41 were obtained, as shown in supplementary Fig.~6. A representative ST-FMR signal ($V_{\rm mix}$) for the 30 nm film is shown in Fig.~\ref{fig. stfmr}b; we did not observe a clear signal for the 20 and 10 nm films. The obtained $V_{\rm mix}$ was fitted to a single Lorentzian function, which is the sum of symmetric and antisymmetric components~\cite{liu2011spin, demasius2016enhanced, behera2022energy}. $\mu_{\rm 0}M_{\rm eff}$, and $\alpha$ values are obtained in the same way as mentioned in the FMR analysis and shown in Fig. \ref{fig. stfmr}c and Fig. \ref{fig. stfmr}d, respectively. The obtained values $\mu_{\rm 0}M_{\rm eff} = 0.79\pm0.04$, and $\alpha=0.018\pm0.001$ are comparable to the values obtained with FMR on blanket films when $H_a$ applied along the (110) plane.

The current dependent ST-FMR measurements were carried out at a fixed frequency to estimate the effective antidamping-like torque efficiency ($\theta_{\rm AD}^{\rm eff}$). The change in $\Delta H$ \textit{vs.} applied dc current, $I_{\rm dc}$, at a frequency of 8 GHz is shown in Fig. \ref{fig. stfmr}e.  
The slope ($\delta \Delta H/\delta I_{\rm dc}$) of linearly fit $\Delta H$ \textit{vs.} $I_{\rm dc}$ from Fig. \ref{fig. stfmr}e indicates the SOT efficiency, which is extracted using ~\cite{liu2011spin, demasius2016enhanced, behera2022energy},

\begin{equation}
   \theta_{\rm AD}^{\rm eff} = \frac{2e}{\hbar}\frac{(H_{\rm a}+0.5M_{\rm eff})\mu_{\rm 0}M_{\rm s}t_{\rm CMG}}{\sin\phi_{\rm H}}\frac{\gamma}{2\pi f}\frac{\delta\Delta H}{\delta I_{\rm dc}}A_{\rm c}
\label{eq3}
\end{equation}
\\
where, $t_{\rm CMG}$ is the thickness of the CMG layer, $\gamma/2\pi$ the effective gyromagnetic ratio, $A_{\rm c}$ the cross-sectional area of the ST-FMR microbars, and $\phi_{\rm H}$ the angle between the applied magnetic field and the rf/dc current. A $\theta_{\rm AD}^{\rm eff}$ value of $1.58\pm 0.40$, and an effective spin Hall conductivity, $\sigma_{\rm SHC}=\sigma_{\rm xx}\theta_{\rm AD}^{\rm eff}$, value of $(1.12\pm 0.30)\times 10^{6}$ ($\hbar/2e$) $(\Omega m)^{-1}$ were estimated for the 30 nm CMG film, corroborating the very high value achieved above using the harmonic Hall measurements.

\begin{figure*}[httb!]
\centering
\includegraphics[width=11.9cm]{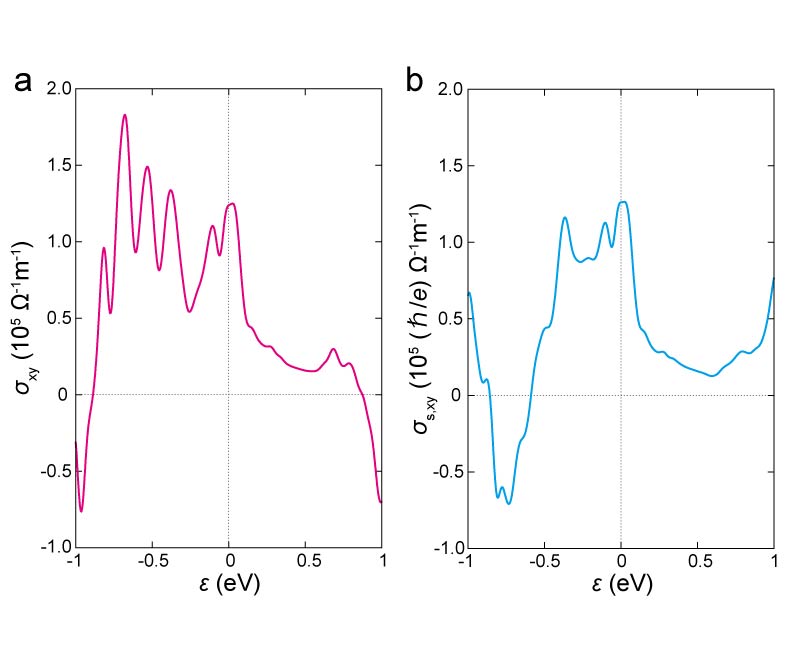}
\caption{\textbf{The physical origin of the giant intrinsic spin Hall conductivity in Co$_{\rm 2}$MnGa.}
\textbf{a}, \textbf{b}, The calculated anomalous Hall and spin Hall conductivities as a function of $\epsilon$ being the energy relative to the Fermi energy.}
\label{fig.theory} 
\end{figure*}

To gain further theoretical insight into the giant spin Hall effect, we calculated the spin Hall conductivity $\sigma_{\rm s,xy}$ in $L$2$_1$-ordered CMG  
by combining first-principles calculation with the Kubo formula. The spin Hall conductivity $\sigma_{\rm s,xy}$ can be calculated in the same way as the anomalous Hall conductivity $\sigma_{\rm xy}$ except that the usual momentum operator $p_y$ is replaced by $p^s_y=\{ p_y,s_z \}$. Fig.~\ref{fig.theory}a shows the energy dependence of the anomalous Hall conductivity $\sigma_{\rm xy}$, where $\epsilon=0$ corresponds to the Fermi level in the present system. We obtained a large value of $\sigma_{\rm xy}$ ($\sim 1.2\times10^5\,\Omega^{-1}{\rm m}^{-1}$) at $\epsilon=0$, consistent with previous studies ~\cite{sakai2018giant,guin2019anomalous,sumida2020spin}. It is known that $L$2$_1$-ordered CMG  
has several mirror symmetries in its crystal structure, and these provide Weyl nodal loops in the $k_i=0$ plane ($i=x,\,y,\,z$) in the Brillouin zone, as shown in the schematic of Fig.~\ref{Fig.1}d ~\cite{guin2019anomalous,manna2018colossal,chang2017topological,sumida2020spin,belopolski2019discovery}. When the spin-orbit interaction is taken into account,
some of these nodal loops are gapped and yield a large value of the Berry curvature, which is the reason for the very large $\sigma_{\rm xy}$ ~\cite{guin2019anomalous,sumida2020spin,manna2018colossal}. In Fig.~\ref{fig.theory}b, we show the energy dependence of the spin Hall conductivity $\sigma_{\rm s,xy}$. We obtained a large $\sigma_{\rm s,xy}$ of $\sim 1.3\times10^5\,(\hbar/e)\,\Omega^{-1}{\rm m}^{-1}$ at $\epsilon=0$, which is comparable to the experimentally obtained large values. Since $\sigma_{\rm s,xy}$ is given as the integral of the spin Berry curvature in the Brillouin zone, the gapped nodal loops are also considered to be the origin of the large spin Hall conductivity. These results clearly indicate that the experimentally obtained large spin Hall effect can be understood by the intrinsic mechanism originating from the electronic structure of $L$2$_1$-ordered CMG.

\begin{figure*}[httb!]
\centering
\includegraphics[width=11.9cm]{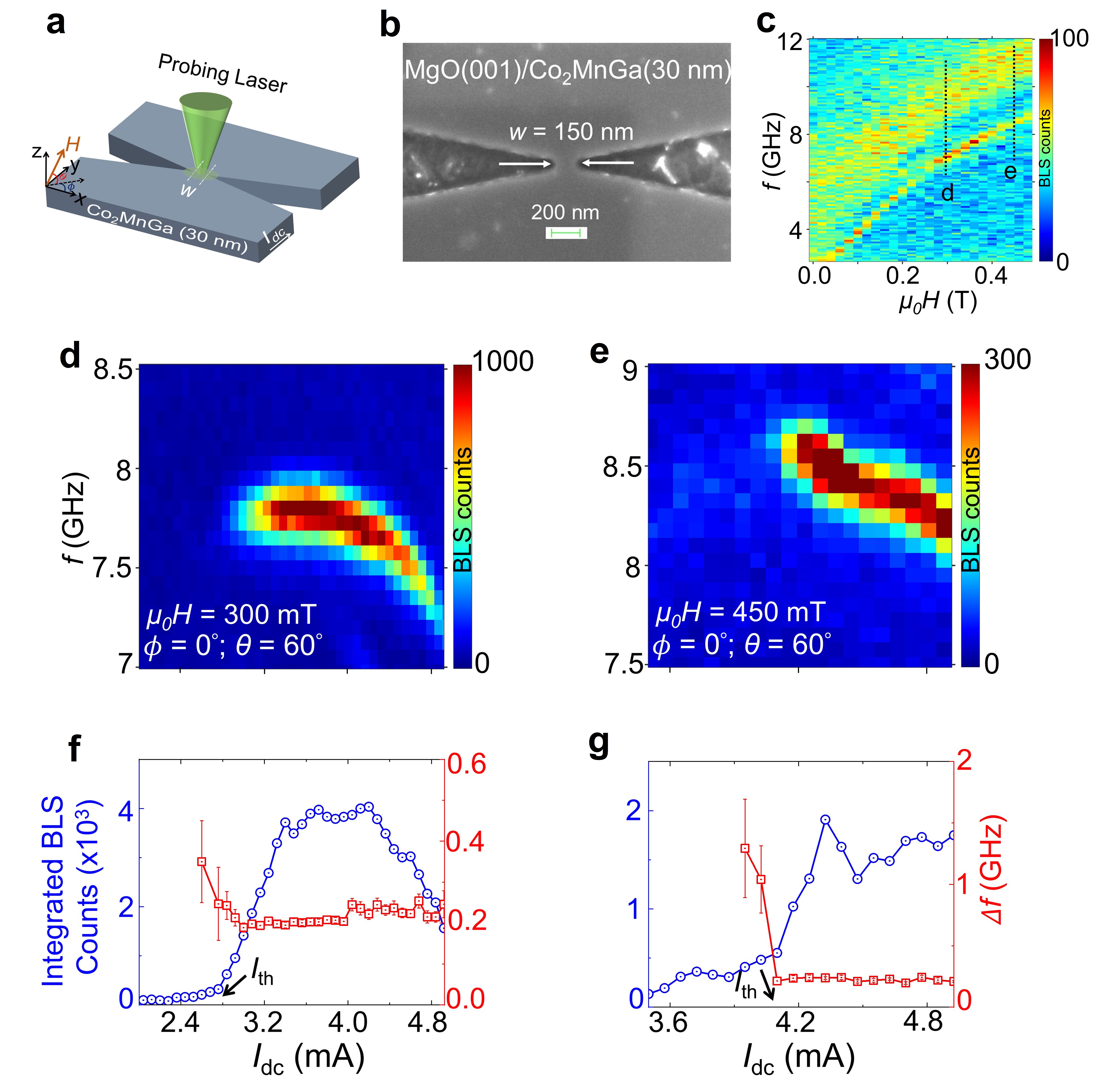}
\caption{\textbf{Magnetization auto-oscillations in 30 nm Co$_{\rm 2}$MnGa.}
\textbf{a}, Schematic of the $\mu$-BLS measurement geometry. \textbf{b}, SEM image of the fabricated 150 nm wide nano-constriction SHNO based on the 30 nm CMG film. \textbf{c}, $\mu$-BLS measurements of the thermal spin wave spectral distribution \emph{vs.}~field strength measured at $I_{\rm dc}=$ 0. \textbf{d,e}, Current dependent auto-oscillation signal measured in a 60$^\circ$ out-of-plane magnetic field $\mu_{\rm 0}H =$ 300 mT, and 450 mT, respectively. \textbf{g, h}, The integrated BLS counts and linewidth as a function of $I_{\rm dc}$ for $\mu_{\rm 0}H =$ 300 mT, and 450 mT, respectively. The threshold current ($I_{th}$) is marked with an arrow in each plot.}
\label{fig.AO} 
\end{figure*}

To confirm and use the observed giant intrinsic SOT,  
we fabricated 150 nm wide nano-constriction SHNOs \cite{demidov2014nanoconstriction,haidar2019single,kumar2022fabrication} out of the 30 nm CMG film and measured their spin wave spectrum as a function of field and SHNO current using 
$\mu$-BLS microscopy. Fig.~\ref{fig.AO}a shows the schematic of the $\mu$-BLS measurements with the SHNO layout, the magnetic field geometry, and the laser spot ($\sim$~300 nm) focused onto the center of the nano-constriction.  
The external magnetic field is applied at an out-of-plane (OOP) angle $\theta =$ 60$^{\circ}$ and an in-plane angle $\phi =$ 0$^{\circ}$. Figure~\ref{fig.AO}b shows a scanning electron microscopy (SEM) image of the SHNO. Additional details of the SHNO device fabrication and $\mu$-BLS measurements are described in the Methods section.  

Figure~\ref{fig.AO}c shows the spectral distribution of the thermal (zero current) BLS counts  \emph{vs.}~field strength. As expected for  
a material without perpendicular magnetic anisotropy (PMA), we observe a wide band of spin waves with a lower FMR cutoff following a Kittel-like frequency-\emph{vs.}-field dependence and a higher more gradual decay as the spin waves approach the wave vector resolution limit of the BLS microscope. Well below the FMR cutoff, we also observe a narrow spin wave mode, which we identify as the typical nano-constriction edge mode.

Figure~\ref{fig.AO}d\&e show the corresponding spin wave spectral distribution \emph{vs.}~SHNO current for two different magnetic fields of 300 mT and 450 mT. The thermal BLS counts are now entirely swamped by spin wave auto-oscillations starting at about 2.8 and 4.1 mA, respectively. The auto-oscillations occur on the nano-constriction edge mode with a frequency that decreases with current, consistent with the negative non-linearity~\cite{dvornik2018origin}.

To confirm the first estimates of the auto-oscillation threshold currents, we fit a Lorentzian to the BLS spectral distribution at each current and plot the integrated BLS counts and the distribution width \emph{vs.}~SHNO current in Fig.~\ref{fig.AO}f\&g. At threshold, the integrated BLS counts show a sharp change of slope and the extracted linewidth drops to the instrument frequency linewidth of the BLS microscope ($\sim$200 MHz). Using these observations as approximate criteria for the threshold current, we extract 2.8 mA at 300 mT and 4.1 mA at 450 mT.  
This translates into threshold current densities of $J_{th}=6.2\times10^{11}$ $Am^{-2}$ and $J_{th}=9.1\times10^{11}$ $Am^{-2}$, which is one order of magnitude lower than 
that of 15 nm single-layer NiFe nano oscillators~\cite{haidar2019single}. Considering that CMG and NiFe have about the same saturation magnetization and that the film in our CMG SHNO is twice as thick as the NiFe in~\cite{haidar2019single}, the ultra-low threshold current density is a strong confirmation of the giant intrinsic SOT of CMG. 

\section*{Conclusion}\label{sec2}
High-quality epitaxial thin films of the ferromagnetic Heusler alloy Co$_{\rm 2}$MnGa were grown with high structural ordering and their intrinsic spin-orbit torque was studied both experimentally and theoretically. 
High, bulk-like, values of the anomalous Hall conductivity, $\sigma_{\rm xy}=1.35\times10^{5}$ $\Omega^{-1} m^{-1}$, and the anomalous Hall angle, $\theta_{\rm H}=15.8\%$ were obtained, confirming 
the high film quality.  
The SOT efficiency was measured using both second harmonic Hall resistance and ST-FMR measurements, and a $\sigma_{\rm SHC}$ value of $(6.08\pm 0.02)\times 10^{5}$ ($\hbar/2e$) $\Omega^{-1} m^{-1}$ was obtained for the 30 nm CMG film, which is an order of magnitude higher than values reported in the literature for any single layer magnets  
and for multilayer Co$_{\rm 2}$MnGa stacks.  
Steady-state magnetization auto-oscillations, at an ultra-low current density of $J_{th}=6.2\times10^{11}$ $Am^{-2}$, were observed using micro-Brillouin light scattering microscopy in nano-constriction SHNOs made from the 30 nm CMG film, confirming its 
giant intrinsic SOT. 
Theoretical calculations explain the origin of the giant intrinsic SOT as due to the large Berry curvature.  
Our study opens up a new research direction to design spintronic devices based on single-layer magnetic Weyl semimetals.

\section*{Methods}\label{sec3}
\subsection*{Sample fabrication}
Epitaxial thin films of CMG (\textit{t} = 10, 20, and 30 nm) are grown on 0.5 mm thick single crystalline MgO(001) substrates using ultra-high vacuum magnetron sputtering with a base pressure of less than $2\times10^{-10}$ torr. All films are deposited at room temperature and followed by post-annealing at $550^\circ$C for 30 minutes. After cooling down the samples to room temperature, a 2 nm thick Al capping layer is deposited to protect the films from oxidation. E-beam lithography, argon milling, and a negative e-beam resist (maN 2401) as an etching mask were then used to fabricate 500 nm wide cross Hall bars, 4\,$\times$14~$\mu m^2$ ST-FMR microstrips, and 150 nm wide nano-constriction SHNOs. 
Optical lithography is used to define the top coplaner waveguide contacts for the electrical measurements, followed by a lift-off process of 780~nm of copper and 20~nm of platinum. The detailed nano-fabrication process can be found in Ref. \cite{kumar2022fabrication}.

\subsection*{Characterization of Co$_{\rm 2}$MnGa (CMG) films}
The composition of the CMG films is determined using X-ray fluorescence spectroscopy. The structural analysis is done using X-ray diffraction (XRD) measurements 
for different CMG atomic planes 
using different tilt angles $\chi$. The longitudinal and anomalous Hall resistivities are measured using a physical property measurement system (PPMS; Quantum Design) at temperatures 50-300 K. Magnetization measurements are done using a vibrating sample magnetometer at room temperature.
Broadband ferromagnetic resonance (FMR)  measurements are done using a NanOsc PhaseFMR-40 system with a co-planar waveguide for broadband microwave field excitation at room temperature. Microwave excitation fields $\mathit{h_{\rm rf}}$ with frequencies up to 38 GHz are applied in the film plane and perpendicular to the applied in-plane dc magnetic field $H_{a}$.

\subsection*{Harmonic Hall measurements}
 
The effective fields of field-like ($H_{\rm FL}$) and antidamping-like ($H_{\rm AD}$) SOTs are evaluated using extended harmonic Hall measurements, 
excluding the thermoelectric effects originating due to the anomalous Nernst and spin Seebeck effects ~\cite{avci2014interplay, takeuchi2018spin}. The schematic of the harmonic Hall measurement setup is shown in Fig. \ref{Fig. 2nd harmonic}a, where a 213 Hz alternating current ($I_{AC}$) is applied to the channel in the presence of a fixed  
magnetic field, $\mu_{\rm 0}H_{\rm a}$. The first and second harmonic Hall voltages ($V_{\omega}$ and $V_{2\omega}$) are measured at room temperature using a lock-in-amplifier while sweeping the in-plane angle, $\phi_{\rm H}$, between the $I_{AC}$ and $\mu_{\rm 0}H_{\rm a}$, as shown in Fig. \ref{Fig. 2nd harmonic}a.      

\subsection*{Magnetoresistance and spin torque ferromagnetic resonance (ST-FMR) measurements}
In-plane angular dependent anisotropic magnetoresistance measurements are performed on 4\,$\times$14 $\mu m^2$ ST-FMR microstrips at room temperature using a rotatable projected vector field magnet with a magnetic field magnitude of 0.1 T and an applied dc current of 0.5 mA. 
Room-temperature ST-FMR measurements are performed by injecting a 
radio-frequency (rf) current 
to the microstrip through a high-frequency bias-T at a fixed frequency (ranging from 7 to 17 GHz) with an input power of $P=$ 4 dBm. The rf current generates antidamping-like and field-like torques in the presence of an applied magnetic field $\mu_{\rm 0}H_{\rm a}$, and the resultant torques excite the magnetization procession of the CMG film, which leads to a time-dependent change in the device resistance due to the magnetoresistance of the CMG~\cite{liu2011spin, demasius2016enhanced}. The oscillating resistance of the device mixes with the rf current and results in a dc mixing voltage, V$_{\rm mix}$, which is then measured using a lock-in-amplifier. ST-FMR measurements are performed with a fixed in-plane angle, $\phi_{\rm H}=60^{\circ}$, between the applied magnetic field and input rf/dc current.  

\subsection*{Micro-Brillouin light scattering ($\mu$BLS) measurement}
Magneto-optical measurements of the SHNOs are carried out using micro-focused Brillouin light scattering microscopy. A monochromatic continuous wave (CW) laser (wavelength = 532 nm; laser power = 1.5 mW) was focused on the center of the nano-constriction by a $\times100$ microscope objective with a large numerical aperture (NA = 0.75) down to a 300 nm diffraction limited spot diameter. The magnetic field condition was set at an in-plane angle (IP) of 0$^{\circ}$, and an out-of-plane (OOP) angle of 60$^{\circ}$.  
The scattered light was 
analyzed with a Sandercock-type six-pass Tandem Fabry-Perot interferometer TFP-1 (from JRS Scientific Instruments). The resulting BLS intensity is proportional to the square of the amplitude of the dynamic magnetization at the location of the laser spot. A special stabilization protocol based on an active feedback algorithm (THATec Innovation) was employed to get long-term spatial stability during the $\mu$-BLS measurements. All the measurements were performed at room temperature. 

\subsection*{Theoretical calculations of anomalous Hall and spin Hall conductivities}
The anomalous Hall and spin Hall conductivities of CMG 
are calculated by combining first-principles calculations and the Kubo formula. First, we calculated the electronic structure of $L$2$_1$-ordered CMG  
(Fig. \ref{Fig.1}a) based on density-functional theory, including the spin-orbit interaction, which is implemented in the Vienna {\it ab initio} simulation program (VASP)~\cite{kresse1996efficient}. The lattice constant is set to a typical experimental value of 5.755\AA~\cite{sumida2020spin}. We adopted the generalized gradient approximation for the exchange-correlation energy and used the projected augmented wave pseudopotential to treat the effect of core electrons properly. A cut-off energy of 337\,eV is employed, and the Brillouin-zone integration is performed with 91\,$\times\,$91$\,\times\,$91 k points. The convergence criteria for energy and force are set to 10$^{-5}$\,eV and 10$^{-4}$\,eV/\AA, respectively. Using the obtained electronic structure, we calculated the anomalous Hall and spin Hall conductivities using the following expressions derived from the Kubo formula:
\begin{eqnarray}
   \sigma_{\rm xy}(\epsilon) &=& -\frac{e^2}{\hbar} \int \frac{d^3k}{(2\pi)^3}\,\, \Omega^{c}_{\rm xy}({\bf k}),\label{eq:AHC}\\
   \sigma_{\rm s,xy}(\epsilon) &=& -\frac{e}{\hbar} \int \frac{d^3k}{(2\pi)^3}\,\, \Omega^{s}_{\rm xy}({\bf k}),\label{eq:SHC}\\
   \Omega^{\alpha}_{\rm xy}({\bf k}) &=& -\frac{\hbar^2}{m^2}\, \sum_{\rm n} f(E_{\rm n,{\bf k}},\epsilon) \sum_{\rm n' \neq n} \frac{2\,{\rm Im} \braket{\psi_{\rm n,{\bf k}}\lvert p_{\rm x} \rvert \psi_{\rm n',{\bf k}}} \braket{\psi_{\rm n',{\bf k}} \lvert p^{\alpha}_{\rm y} \rvert \psi_{\rm n,{\bf k}}}}{(E_{\rm n',{\bf k}}-E_{\rm n,{\bf k}})^2},\label{eq:BC}
\end{eqnarray}
where $\sigma_{\rm xy}(\epsilon)$ and $\sigma_{\rm s,xy}(\epsilon)$ are the anomalous Hall and spin Hall conductivities, respectively, as a function of $\epsilon$ being the energy relative to the Fermi energy. These conductivities are given by integrating the charge Berry curvature $\Omega^{c}_{\rm xy}$ and the spin Berry curvature $\Omega^{s}_{\rm xy}$, where the generalized momentum operator $p^\alpha_y$ is defined as $p^c_y=p_y$ and $p^s_y=\{ p_y,s_z \}$ with the spin operator $s_z=\sigma_z/2$~\cite{tanaka2008intrinsic}. In Eq.~(\ref{eq:BC}), $\lvert\psi_{\rm n,{\bf k}}\rangle$ is the eigenstate with the eigenenergy $E_{\rm n,{\bf k}}$ for the band $n$ and the wave vector ${\bf k}$, and $f(E_{\rm n,{\bf k}},\epsilon)$ is the Fermi distribution function. In our calculations, the direction of the magnetization is fixed to the [001] direction, consistent  
with the experimental setup.

\section*{Declarations}

\bmhead{Funding}
Lakhan Bainsla thanks MSCA - European Commission for the Marie Curie Individual Fellowship (MSCA-IF Grant No. 896307) and the Science and Engineering Research Board (SERB) India for the Ramanujan fellowship. This work was partially supported by the Swedish Research Council (VR Grant No. 2016-05980), the Horizon 2020 research and innovation programme (ERC Advanced Grant No.~835068 "TOPSPIN"), FLAG-ERA project 2DSOTECH (VR No. 2021-05925), and EU project 2DSSPIN-TECH.

\bmhead{Conflict of interest}
The authors declare no competing interests.

\bmhead{Availability of data and materials}
The data is available upon reasonable request from corresponding authors.

\bmhead{Authors' contributions} 
L.B., Y.S., S.P.D. and J.Å. conceived the idea and planned the study. Y.S. prepared the thin film samples and performed the structural, magnetic, and transport measurements on the thin film samples. L.B. fabricated the devices and performed the FMR, harmonic Hall, and ST-FMR measurements and analysis. A.K.C. and A.A.A. performed all micro-BLS measurements and analysis. A.K. recorded the scanning electron microscopy images of SHNOs. A.K. and S.P.D. supported L.B. in the harmonic Hall data analysis and preparation of figures. K.M. did all the theoretical calculations. J.Å. coordinated and supervised the study. All authors discussed the results and co-wrote the manuscript.

\bibliography{references}

\end{document}